\documentclass[showpacs,aps,prb,reprint]{revtex4-1}
\usepackage{graphicx}
\usepackage{amssymb}
\usepackage{amsmath}
\usepackage[utf8]{inputenc}
\usepackage{bm}
\usepackage{color}

\newcommand{\bcfsw}[1]{B^{(\textit{\tiny\textsc{FSW, FPv}})}_{c,#1}}

\begin{document}
\title{Magnetic field - temperature phase diagram of ferrimagnetic alternating chains: spin-wave theory from a fully polarized vacuum}
\author{W. M. da Silva}
\author{R. R. Montenegro-Filho}
\affiliation{Laborat\'{o}rio de F\'{i}sica Te\'{o}rica e Computacional, Departamento de F\'{i}sica, 
Universidade Federal de Pernambuco, 50760-901 Recife-PE, Brasil}
\date{\today}

\begin{abstract}
Quantum critical (QC) phenomena can be accessed by studying quantum magnets under an applied magnetic field ($B$).
The QC points are located at the endpoints of magnetization plateaus and separate gapped and gapless phases. 
In one dimension, the low-energy excitations of the gapless phase form a Luttinger liquid (LL), and crossover lines bound 
insulating (plateau) and LL regimes, as well as the QC regime. Alternating ferrimagnetic chains have a spontaneous magnetization 
at $T=0$ and gapped excitations at zero field. Besides the plateau at the fully polarized (FP) magnetization; due to the gap, 
there is another magnetization plateau at the ferrimagnetic (FRI) magnetization. We develop spin-wave theories to study the thermal 
properties of these chains under an applied magnetic field: one from the FRI classical state, and other from the FP state, comparing their 
results with quantum Monte Carlo data. We deepen the theory from the FP state, obtaining the crossover lines in the $T$ vs. $B$ low-$T$ phase 
diagram. In particular, from local extreme points in the susceptibility and magnetization curves, we identify the crossover between an LL regime 
formed by excitations from the FRI state to another built from excitations of the FP state. These two LL regimes are bounded by an asymmetric 
dome-like crossover line, as observed in the phase diagram of other quantum magnets under an applied magnetic field.
\end{abstract}

\maketitle

\section{Introduction}

The theory of quantum phase transitions \cite{sachdev2001quantum,Vojta2003} 
provides a framework from which the low-temperature behavior of many
condensed-matter systems can be understood. 
The quantum critical 
point separates an insulating gapped phase and a gapless conducting phase.
Of particular importance are
magnetic insulators \cite{Zapf2014,Giamarchi2008}, for which the quantum critical regime 
can be experimentally accessed through an applied magnetic field. In these systems, 
the gapped phases are associated to magnetization plateaus in the magnetization curves.

In one dimension, magnetization plateaus can be 
understood as a topological effect through the 
Oshikawa, Yamanaka, and Affleck (OYA) argument \cite{PhysRevLett.78.1984}, which generalizes 
the Lieb-Schultz-Mattis theorem \cite{Lieb1961}.
The OYA argument asserts that a magnetization plateau is possible only if $(S_u-m_u)=\text{integer}$, where 
$m_u$ is the ground-state magnetization and $S_u$ is the sum of the spins in a unit period of the ground state, respectively.
If the ground state does not present spontaneous translation symmetry breaking, $S_u$ is equal to the 
fully polarized magnetization per unit cell, while $m_u$ is the magnetization per unit cell of the system.
The OYA argument was further extended \cite{Oshikawa1999} to models in higher dimensions and to charge degrees of freedom.

Due to the gap closing a magnon excitation, the endpoints of magnetization plateaus are quantum critical points. 
In three-dimensional systems, this transition is 
in the same universality class of the Bose-Einstein condensation \cite{Giamarchi2008,PhysRevB.43.3215}
and was studied in a variety of magnetic insulators \cite{Giamarchi2008,Zapf2014,Paduan-Filho2012}.
In the magnetic system, the magnetization and the magnetic field play the role of the boson density and of the chemical potential, respectively, 
of the bosonic model.
In one dimension the mapping to a hard-core boson model or a spinless fermion system \cite{PhysRevB.43.3215} 
implies a square-root singularity in the magnetization curve: $m\sim \sqrt{|B-B_c|}$ as $B\rightarrow B_c$; 
and, if three-dimensional couplings are present, the condensate can be stabilized at temperatures below
that of the three-dimensional ordering \cite{PhysRevB.43.3215}.

Exactly at the quantum critical field, the magnons have a classical dispersion relation,  
$\omega\sim q^2$, where $q$ is the lattice wave-vector. In one dimension, this quantum critical field separates a gapped phase 
from a gapless Luttinger liquid (LL)
phase \cite{giamarchi2003quantum,GIAMARCHI2012}, with excitations showing a linear dispersion relation, $\omega\sim q$.
The predictions of the Luttinger liquid theory in magnetic insulators with a magnetic field, including the quantum critical regime,  
were investigated in many materials \cite{Ward2017,e247202,PhysRevB.83.054407}.
For finite temperatures and $B\approx B_c$, the quantum critical regime is observed, and the crossover 
line \cite{PhysRevLett.99.057205} to the LL regime is given by $T(B)\sim a|B-B_c|$,
with a universal, model-independent, coefficient $a$.  

One-dimensional ferrimagnets \cite{JBCS2008,PhysRevB.29.5144} show spontaneous 
magnetization at $T=0$, as expected from the Lieb and Mattis theorem \cite{Lieb.Mattis}, 
and a gap in the excitation spectrum is responsible for a magnetization plateau in 
their magnetization curves at the ground-state magnetization value.
In zero field, the critical properties in the vicinity of the thermal critical point at $T=0$ were studied 
in the isotropic \cite{PhysRevLett.78.4853,*PhysRevB.59.14384,AlcarazandMa}
and anisotropic cases \cite{AlcarazandMa}.
Interesting physics emerges through the introduction of destabilizing factors 
of the ferrimagnetic state, such as
doping \cite{PhysRevLett.74.1851,RenePRB2006,PhysRevB.59.7973,Rojas2012,Lopes2014,Montenegro-Filho2014,Kobayashi2016} or 
geometric frustration \cite{Hida1994,Takano1996,RenePRB2008,Ivanovart10,Shimokawa2011,Furuya2014,Amiri2015,StreckaPRB2017,Hida2017}. 
The spin-wave theory \cite{Noriki2017} of ferrimagnetic chains
\cite{PatiJPCM1997,*PhysRevB.55.8894,
Brehmer1997,PhysRevB.57.R14008,PhysRevB.57.13610,Maisinger1998,JCPM.10.11033.1998,Ivanov2000,PhysRevB.69.06,Noriki2017} 
was developed from the classical ferrimagnetic ground state, 
considering free and interacting magnons, with emphasis on zero-field properties. 
The magnetization curves of these systems under an applied magnetic field were  
discussed mainly through numerical methods \cite{PatiJPCM1997,*PhysRevB.55.8894,Maisinger1998,Gu2006,PhysRevB.80.014413,Gong2010,ReneJPCM2011,Strecka2017,StreckaActa2017}.

In this work, we investigate the spin-wave theory of ferrimagnetic alternating chains at 
low temperatures and in the presence of a magnetic field. We compare some results with 
quantum Monte Carlo (QMC) data, obtained using the stochastic series expansion method code from the Algorithms and Libraries for 
Physics Simulations (ALPS) project \cite{Bauer2011}, with $1\times 10^6$ Monte Carlo steps.
We consider spin-wave excitations from the ferrimagnetic and fully polarized classical states. 
In the ferrimagnetic case, we consider interacting spin-waves, while in the 
fully polarized, only free spin-waves are discussed. Considering the whole values of magnetization, 
from zero to saturation, the two approaches present similar deviations from 
the QMC data. We deepen the theory from the ferromagnetic ground state and obtain the 
crossover lines bounding the plateau and LL regimes. In particular, we show that susceptibility and magnetization 
data can be used to identify a crossover between two LL regimes, one built from excitations of the ferrimagnetic 
magnetic state, and the other from the fully polarized one.

This paper is organized as follows. In Sec. \ref{sec:model-qmc} we present the Hamiltonian model and discuss 
the magnetization curves from QMC calculations. In Sec. \ref{sec:sw-theory}
the spin-wave theories from the FRI and FP classical states are discussed, 
particularly the methodology used to obtain the respective 
magnetization curves with a finite temperature, and  make  
a comparison between their results and QMC data. In Sec. \ref{sec:ll-regime}, 
we study LL and plateau regimes at finite temperature through the free spin-wave (FSW) theory from 
the FP vacuum (FSW-FPv). Finally, in Sec. \ref{sec:summary-pd} we summarize our results 
and sketch the $T$-$B$ phase diagram from the FSW-FPv theory of the alternating (1/2,1) spin chain.

\section{Model Hamiltonian and QMC magnetization curves}
\label{sec:model-qmc}

An alternating spin ($s$, $S$) chain has two kinds of spin, $S$ and $s$, alternating on a ring with antiferromagnetic superexchange coupling $J$ 
between nearest neighbors, and described by the Hamiltonian
\begin{equation}
\mathcal{H}=J\sum_{j=1}^N\Big(\mathbf{s}_{j}\cdot\mathbf{S}_{j}  +  \mathbf{s}_{j}\cdot \mathbf{S}_{j+1}\Big)-B\sum_{j}^N(S_{j}^{z} + s_{j}^{z}),
\label{HeisFerri}
\end{equation}
where $B$ is the magnetic field and $N$ denotes the number of unit cells.  We assume $S>s$ and consider equal
$g$-factors for all spins, defining $g\mu_B=1$, where 
$\mu_B$ is the Bohr magneton. The magnetization per unit cell is given by
\begin{equation}
 m = \sum_{j}^{N}(S_{j}^{z} + s_{j}^{z}).
\end{equation}
\begin{figure}[!htb]
\begin{center}
\includegraphics[width=0.4\textwidth]{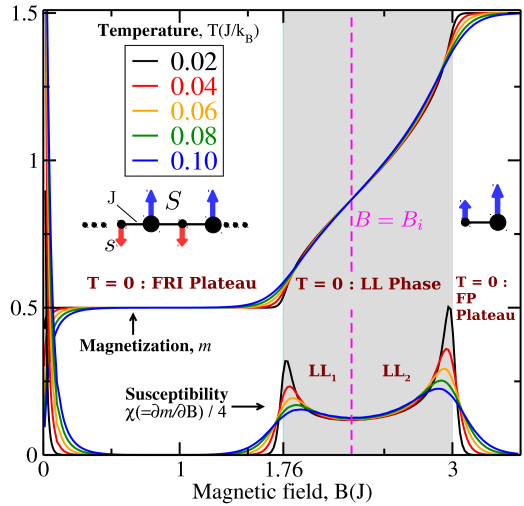}
\caption{(color online). Magnetization plateaus at finite temperature, Luttinger liquid phase and crossovers: Quantum Monte Carlo (QMC) data. 
Magnetization per cell $m$ and the susceptibility $\chi=\partial m/\partial B$ as a function of magnetic field $B$ for an 
alternating ($s=1/2$, $S=1$) chain with $N=256$ unit cells and the indicated values of temperature $T$. 
The critical endpoint of the ferrimagnetic (FRI) and the fully polarized (FP) plateaus are $B_{c, FRI}=1.76J$ and
$B_{c,FP}=3J$, respectively. The presence of the FRI and FP plateaus, and the 
region dominated by Luttinger liquid (LL) regime is a common feature 
for all values of $s$ and $S$, with $S>s$. 
As $T\rightarrow 0$, $\chi\rightarrow\infty$ at the critical values of $B$; for $T\gtrsim0$, local maxima 
in the $\chi$ curves marks the crossover from the LL regime to the quantum critical regime. The local minimum 
in the $\chi$ curve (dashed line) between $B_{c,FRI}$ and $B_{c,FP}$ separates the LL regime into two regions: one 
with excitations from the FRI state, $\text{LL}_1$; the other with excitations from the FP state, $\text{LL}_2$.
}
\label{fig:mb0-01}
\end{center}
\end{figure}

In Fig. \ref{fig:mb0-01} we show QMC results for $m(B)$ for the (1/2, 1) chain in the low-$T$ regime.
At $T=0$, $m(B)$ presents two magnetization plateaus: the ferrimagnetic (FRI), at $m_{FRI}=(S-s)$, and the fully polarized (FP) one, 
at $m_{FP}=s+S$. In particular, at $T=0$, $m=m_{FRI}$ for $B=0$, with a gapless Goldstone mode. There are quantum phase transitions at the endpoint of the plateaus: $B=B_{c,FRI}$ and 
$B=B_{c,FP}$, respectively; which have the values $B_{c,FRI}=1.76J$ and $B_{c,FP}=3.00J$ for the ($1/2$, $1$) chain. 
At the critical fields, there is a transition 
from a gapped plateau phase to a gapless Luttinger liquid (LL) phase,
as $B\rightarrow B_{c,FRI}$ from magnetic fields $B<B_{c,FRI}$, or $B\rightarrow B_{c,FP}$ from magnetic fields $B>B_{c,FP}$.
In the LL phase, the excitations have 
a linear dispersion relation, $\omega\sim q$, and present critical (power-law) transverse spin correlations.
Exactly at the critical fields, the excitations have a classical dispersion relation $\omega\sim q^2$ and 
in the high diluted limit can be represented by a hard-core boson model or a spinless fermion model.
Hence, the magnetization has a square-root behavior $m\sim\sqrt{|B-B_c|}$ and a diverging susceptibility
$\chi=\partial m/\partial B\sim 1/\sqrt{|B-B_c|}$ as $B\rightarrow B_{c}$. 

For finite-$T$, but $T\rightarrow0$, the magnetization $m=0$ for $B=0$, since the system is one-dimensional. 
Gapped magnetic excitations are thermally activated and the plateau widths reduce.
The susceptibility shows local maxima, with distinct amplitudes, 
at $B\approx B_{c,FRI}$ and $B\approx B_{c,FP}$ marking the crossover between the LL regime, where the excitations 
have a linear behavior, $\omega\sim q$, to the quantum critical regime, for which $\omega\sim q^2$.
We can define the local minimum in the $\chi$ curve, at $B\equiv B_i$, 
as a crossover between the region where the excitations are predominantly from the FRI state, denoted by LL$_1$ in Fig. \ref{fig:mb0-01}, 
and that where the excitations are predominantly from the FP state, denoted by LL$_2$ in Fig. \ref{fig:mb0-01}. In particular, 
for $B\approx B_i$, the magnetization curve has its more robust value and behavior as the temperature increases, showing 
that the LL phase is more robust for $B\approx B_i$.
\section{Spin-wave Theory}
\label{sec:sw-theory}

The ferrimagnetic arrangement of classical spins is a natural choice of vacuum to study quantum ferrimagnets through free 
spin-wave (FSW) theory \cite{PatiJPCM1997,*PhysRevB.55.8894}, if we want to study excitations from the quantum ground state. Two types of magnon excitations are obtained, one ferromagnetic, which decreases the ground state spin 
by one unit, and the other antiferromagnetic, increasing the ground state spin by one unit. In particular, the antiferromagnetic 
excitation has a finite gap $\Delta$, which implies the expected magnetization plateau at $m=S-s$ and $T=0$.
However, at this linear approximation, quantum fluctuations are underestimated, giving poor 
results for the value of antiferromagnetic gap, 
and other quantities, like the average spin per site.  

When one-dimensional ferromagnets are studied through the linear spin-wave theory at finite temperatures, a diverging zero-field magnetization 
is obtained for any value of $T$\cite{PhysRev.58.1098, PhysRev.86.694, PhysRev.87.568}.
Takahashi \cite{Prog.Theor.Phys.Supp.87.233, PhysRevLett.58.168} modified the theory by imposing a constraint
on the zero-field magnetization and an effective chemical potential in the thermal boson distribution.
This so-called modified spin-wave theory describes very well the low-temperature thermodynamics of one-dimensional 
ferromagnets, and was further successfully adapted to other systems, including ferrimagnetic chains \cite{PhysRevB.57.R14008}.
In the case of ferrimagnets, the introduction of the magnetization constraint in the bosonic distribution, with the 
linear spin-wave dispersion relations gives an excellent description of the low-$T$ behavior.
The description of the intermediate-$T$ regime can be improved by changing the 
constraint \cite{Noriki2017}. 

In this Section, we discuss interacting spin-wave theory using a ferrimagnetic vacuum (ISW-FRIv) for $B\neq0$ and $T\neq0$, 
with the modified spin-wave approach (Takahashi's constraint);
and free spin-wave theory from a fully polarized vacuum (FSW-FPv), also for $B\neq0$ and $T\neq0$.

\subsection{Spin-wave theory - ferrimagnetic vacuum}
\label{Section_FRIv}
\begin{figure}[!htb]
\includegraphics[width=0.47\textwidth]{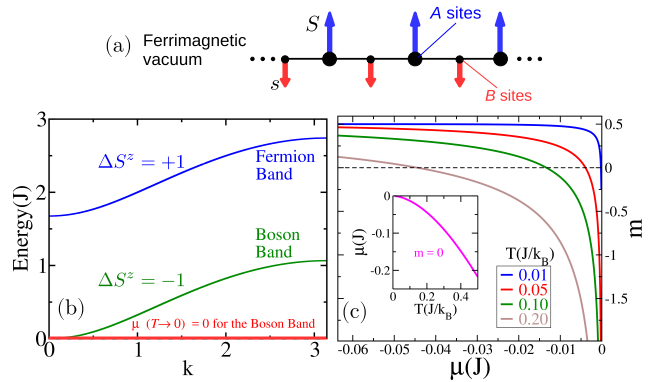}
\caption{(color online). Interacting spin-wave (ISW) magnon branches from the classical ferrimagnetic vacuum (FRIv) - calculating the thermodynamic properties.
(a) The classical ferrimagnetic vacuum of the ($s$,$S$) chain.
(b) Magnon dispersion relations
for the ($s=1/2$, $S=1$) chain with $B=0$.
There are ferromagnetic and antiferromagnetic 
magnons, carrying spin $\Delta S^z=-1$ and $\Delta S^z=1$, respectively. The values of the critical fields
are $B^{(\textsc{ISW-FRIv})}_{c,\textsl{FRI}}=1.68J$ and 
$B^{(\textsc{ISW-FRIv})}_{c,\textsl{FP}}=2.74J$.
To calculate the thermodynamic functions, the antiferromagnetic (ferromagnetic) magnons occupies 
their respective bands following the Fermi (Bose) distribution function.
An effective chemical potential $\mu$ is introduced in the Bose distribution to prevent particle 
condensation at the $k=0$ mode for $B=0$ and $T\rightarrow0$.  
(c) For each value of $T$, we use a value of $\mu$ such that $m=0$ for $B=0$.
The inset shows that $\mu(T\rightarrow0)\rightarrow0$ as $T\rightarrow0$. 
In this limit, both bands are empty and $m=(S-s) =1/2$, the FRI magnetization. 
}
\label{fig:swfrit}
\end{figure}
The Holstein-Primakoff spin-wave theory is developed from the classical ground state illustrated in Fig. \ref{fig:swfrit}(a), 
which has the energy $E^{\text{\tiny\textsc{(FRIv)}}}_{class} = -2JNsS -B\big(S-s\big)N$.
The bosonic operators $a_j$ ($a^\dagger_j$) and $b_j$ ($b^\dagger_j$), associated to $A$ and $B$ sites, respectively, 
have the following relation with the spin operators (Holstein-Primakoff transformation):

\begin{eqnarray}
S^+_j &=&\sqrt{2S}\Big(1-\frac{a^\dagger_j a_j}{2S}\Big)^{1/2}a_j\text{, and }S^z_j=S-a^\dagger_j a_j;\\ 
s^+_j &=&b^{\dagger}_j\sqrt{2s}\Big(1-\frac{b^\dagger_j b_j}{2s}\Big)^{1/2}\text{, and }s^z_j=b^\dagger_j b_j-s.
\label{eq:hp-relations}
\end{eqnarray}

Putting the Hamiltonian (\ref{HeisFerri}) in terms of these bosonic operators, expanding
to quadratic order, Fourier transforming and making the following Bogoliubov transformation \cite{PatiJPCM1997,*PhysRevB.55.8894}: 
\begin{equation*}
 a_{k} = \alpha_{k}\cosh\theta_{k} - \beta_{k}^\dag\sinh\theta_{k} ,
 \label{TransfBogol}
\end{equation*}
\begin{equation}
 b_{k} = \beta_{k}\cosh\theta_{k} - \alpha_{k}^\dag\sinh\theta_{k},
\end{equation}
\begin{equation}
 \tanh2\theta_k = 2\frac{\sqrt{sS}}{s+S}\cos \Big(\frac{k}{2}\Big),
 \label{TanBogol}
\end{equation}
where $k$ is the lattice wave-vector, the 
non-interacting spin-wave Hamiltonian is given by
\begin{equation}
 \mathcal{H}^{\text{\tiny\textsc{(FSW-FRIv)}}} = E_0+\sum_{k}\Big[\omega^{\text{\tiny\textsc{(FRIv)}}}_{k,-}\alpha_{k}^\dag \alpha_{k} + \omega^{\text{\tiny\textsc{(FRIv)}}}_{k,+}\beta_{k}^\dag \beta_{k}\Big].
 \label{HeisFerri0Diag}
\end{equation}
The magnon branches obtained are:
\begin{equation}
 \omega^{\text{\tiny\textsc{(FRIv)}}}_{k,\sigma} = \sigma J\big(S-s\big) -\sigma B + J\omega^{\text{\tiny\textsc{(FRIv)}}}_{k},
 \label{RelDispFerriMaisMenos}
\end{equation}
with $\sigma=\pm$, and 
\begin{equation}
 \omega^{\text{\tiny\textsc{(FRIv)}}}_{k} = \sqrt{\big(S-s\big)^2 + 4sS{\sin}^2 \Big(\frac{k}{2}\Big)},
 \label{omegaK}
\end{equation}
while the ground-state energy is
\begin{equation}
E_0 = J\sum_{k}\Big[\omega^{\text{\tiny\textsc{(FRIv)}}}_{k} - \big(S+s\big)\Big].
\end{equation}

The $\omega^{\text{\tiny\textsc{(FRIv)}}}_{k,-}$ modes carry a spin $\Delta S^z=-1$, having a ferromagnetic spin-wave nature, 
and is gapless for $B=0$; while $\omega^{\text{\tiny\textsc{(FRIv)}}}_{k,+}$ modes carry a spin $\Delta S^z=+1$, having an 
antiferromagnetic spin-wave nature and 
has a gap $\Delta=2J(S-s)$ at $B=0$. For the ($s=1/2$, $S=1$) chain \cite{PatiJPCM1997,*PhysRevB.55.8894}, for example, 
$\Delta=1$, 
although the exact value is $1.76J$; while $\langle S^z_a\rangle=0.695$ and
$\langle S^z_b\rangle=-0.195$  at $T=0$, with the exact values \cite{PatiJPCM1997,*PhysRevB.55.8894}: 
$\langle S^z_a\rangle=0.792$ and $\langle S^z_b\rangle=-0.292$.

The dispersion relations can be improved if interactions between magnons are considered.
The corrected dispersion relations described in Ref. \cite{JCPM.10.11033.1998}, shown in Fig. \ref{fig:swfrit}(b), are:
\begin{equation}
 \tilde{\omega}^{\text{\tiny\textsc{(FRIv)}}}_{k,\sigma} = {\omega}^{\text{\tiny\textsc{(FRIv)}}}_{k,\sigma} - J\delta{\omega}^{\text{\tiny\textsc{(FRIv)}}}_{k,\sigma},
 \label{RelDispMelhorada}
\end{equation}
where
\begin{equation*}
\delta \omega^{\text{\tiny\textsc{(FRIv)}}}_{k,\sigma} = 2\Gamma_{1}\frac{(S + s)}{\omega^{\text{\tiny\textsc{(FRIv)}}}_{k}}\sin^{2}(k/2) - \frac{\Gamma_2}{\sqrt{sS}}\Big[\omega^{\text{\tiny\textsc{(FRIv)}}}_k +\sigma (S - s)\Big],
  \label{RelacaoDispersaoMagAntiFerroFerroMelhoradas}
\end{equation*}
with
\begin{eqnarray}
 \Gamma_1 &=& \frac{1}{N}\sum_{k}\sinh^{2} \theta_{k},\text{ and}\\
 \Gamma_2 &=& \frac{1}{N}\sum_{k}\cos(k/2)\sinh \theta_{k}\cosh \theta_{k}.
\end{eqnarray}
Up to $\mathcal{O}(S^0)$, the Hamiltonian is
\begin{equation}
 \mathcal{H}^{\text{\tiny\textsc{(ISW-FRIv)}}}= E_g + \sum_{k}\big(\tilde{\omega}^{\text{\tiny\textsc{(FRIv)}}}_{k,-} \alpha^{\dag}_{k}\alpha_{k} + \tilde{\omega}^{\text{\tiny\textsc{(FRIv)}}}_{k,+}\beta^{\dag}_{k}\beta_{k}\big),
 \label{HamiltonianoFinal}
\end{equation}
where
\begin{equation}
  E_g = E_{class} + E_0 + E_1,
  \label{GScorrigido}
\end{equation}
with
\begin{equation}
 E_1 = -2JN\Big[\Gamma_{1}^{2} + \Gamma_{2}^{2} - \Big(\sqrt{S/s} + \sqrt{s/S}\Big)\Gamma_1\Gamma_2 \Big].
\end{equation}

At $T=0$, the magnetization as a function of $B$, shown in Fig. \ref{fig:mb0-01} for the ($s=1/2$, $S=1$) chain, can be understood from these 
ferromagnetic ($\Delta S^z=-1$) and antiferromagnetic ($\Delta S^z=+1$) magnon modes.
For $B=0$ the two bands are empty and the magnetization is the ferrimagnetic one. Increasing the magnetic field, 
the ferromagnetic band acquires a gap which increases linearly with $B$, while the gap to the antiferromagnetic band
decreases linearly with $B$. Notice, in particular, that the ferromagnetic band is empty for all values 
of $B$. At $B=B^{(\textsc{ISW-FRIv})}_{c,\textsl{FRI}}/2=\Delta/2$, the $k=0$ mode of the antiferromagnetic
band is the lower energy state, and at $B=B^{(\textsc{ISW-FRIv})}_{c,\textsl{FRI}}=\Delta$   
the gap to this mode closes. The value of $B^{(\textsc{ISW-FRIv})}_{c,\textsl{FRI}}$ is
\begin{equation}
B^{(\textsc{ISW-FRIv})}_{c,\textsl{FRI}} =\tilde{\omega}^{\text{\tiny\textsc{(FRIv)}}}_{0,+}= 2(S-s)\left(1 + \frac{1}{\sqrt{sS}}\Gamma_2\right)J.
\end{equation}
In particular, for the ($s=1/2$, $S=1$) chain, with $\Gamma_1=0.305$ and $\Gamma_2=0.478$,  
$B^{(\textsc{ISW-FRIv})}_{c,\textsl{FRI}}=1.68J$, which is very close to the exact value ($1.76J$). 

The magnetization for $B>\Delta$ is obtained by considering the antiferromagnetic magnons as 
hard-core bosons \cite{PhysRevB.43.3215}, or spinless fermions. The magnetization increases with $B$ as the antiferromagnetic band is filled, and 
saturates when the Fermi level reaches the band limit, at $k=\pi$. The saturation field is
\begin{equation}
B^{(\textsc{ISW-FRIv})}_{c,\textsl{FP}}=\tilde{\omega}^{\text{\tiny\textsc{(FRIv)}}}_{\pi,+} = 2\left(S - \Gamma_1+\sqrt{\frac{S}{s}}\Gamma_2\right)J,
\end{equation}
which for the ($s=1/2$, $S=1$) chain is $B^{(\textsc{ISW-FRIv})}_{c,\textsl{FP}}=2.74$, departing from the exact value $3J$, but much better 
than the free spin wave result: $2J$.  

\subsubsection{Thermodynamics}

For $T>0$, ferromagnetic and antiferromagnetic modes are occupied in accord to Bose-Einstein ($n^{\text{\tiny\textsc{(FRIv)}}}_{k,-}$) and 
Fermi-Dirac ($n^{\text{\tiny\textsc{(FRIv)}}}_{k,+}$) distributions, respectively, as indicated in Fig. \ref{fig:swfrit}(a). The magnetization, for example, is given by
\begin{equation}
 m(T,B)= (S - s)+\frac{1}{N}\sum_{k}(n^{\text{\tiny\textsc{(FRIv)}}}_{k,+}-n^{\text{\tiny\textsc{(FRIv)}}}_{k,-}).
 \label{RestricMagnet}
\end{equation}
We notice, however, that with $T>0$ and $B=0$ the ferromagnetic band will be thermally activated and $m\rightarrow-\infty$ as $T$ increases. 
This problem arises, also, in one-dimensional ferromagnetic 
chains, and was overcome by Takahashi \cite{Prog.Theor.Phys.Supp.87.233, PhysRevB.40.2494}, in the low-$T$ regime, through the introduction of
an effective chemical potential $\mu$ in the bosonic distribution, and a constraint $m(B=0,T)=0$. 
A similar strategy was applied to one-dimensional ferrimagnetic systems \cite{PhysRevB.57.R14008}
and good results were also obtained in the low-$T$ regime.
The intermediate-$T$ regime, where the minimum in the $T\chi$ curve of the ferrimagnets \cite{PhysRevB.29.5144} are observed, 
can be more accurately described if other constraints are used
\cite{PhysRevB.69.06,JCPM.10.11033.1998,Noriki2017}.

Here, for $B=0$, we use the simplest constraint 
\begin{equation}
 m(T,B=0)=0,
\end{equation}
since we are interested in the low-$T$ regime, with
\begin{eqnarray}
  n^{\text{\tiny\textsc{(FRIv)}}}_{k,-} &=& \frac{1}{e^{\beta [\tilde{\omega}^{\text{\tiny\textsc{(FRIv)}}}_{k,-} - \mu]} -1},\label{NumMagnons1}\\ 
  n^{\text{\tiny\textsc{(FRIv)}}}_{k,+} &=& \frac{1}{e^{\beta\tilde{\omega}^{\text{\tiny\textsc{(FRIv)}}}_{k,+}} +1}.
  \label{NumMagnons2}
\end{eqnarray}
In Fig. \ref{fig:swfrit}(b), we present $m(T,B=0)$ for the indicated values of $T$. As discussed, $m\rightarrow-\infty$ at $\mu=0$ 
and the value of $\mu$ for which the constraint $m(T,B=0)=0$ is satisfied, monotonically decreases with $T$, in this low-$T$ regime.
A finite $\mu$ implies an effective gap for the ferromagnetic band, with an exponential thermal activation of their magnons. 
In particular, notice that $\mu(T\rightarrow0)=0$, as expected. To calculate the thermodynamic functions for $B\neq0$,
we consider the distributions in Eqs. (\ref{NumMagnons1}) and (\ref{NumMagnons2}) and 
use the same value of $\mu$ found in the case $B=0$: $\mu(B,T)=\mu(B=0,T)$, for any value of $B$. 

The magnetization as a function of $B$ for $T\neq0$, shown in Fig. \ref{fig:mb0-01}, can be qualitatively understood from
this theory. For $B=0$, the magnetization $m=0$, due to the constraint. As $B$ increases, in the region $0<B<B_{c,FRI}/2$, the
gap to the ferromagnetic band increases, but this band is thermally activated and the magnetization decreases from the $m=S-s$ value. 
This effect can also be seen from Fig. \ref{fig:swfrit}(b). If we move the Zeeman term, $+B$, from the ferromagnetic dispersion relation 
to the chemical potential, $\tilde{\omega}^{\text{\tiny\textsc{(FRIv)}}}_{k,-}\rightarrow\tilde{\omega}^{\text{\tiny\textsc{(FRIv)}}}_{k,-}-B$ and $-\mu\rightarrow-(\mu-B)$, in Eq. (\ref{NumMagnons1}), the magnetization value is the one shown in Fig. \ref{fig:swfrit}(b) for 
$\mu$ lower than that of $B=0$, and $m=0$. From Fig. \ref{fig:swfrit}(b), we see that increasing $B$ (decreasing $\mu$) from $B=0$ [from $\mu(B=0,T)$], 
the magnetization rises exponentially to the ferrimagnetic value. 
For $B=B^{(\textit{\tiny\textsc{FSW, FPv}})}_{c,FRI}/2$, the lower energy band is the antiferromagnetic ($\Delta S^z=+1$ magnons) fermionic band. 
This band is thermally activated for $[B^{(\textit{\tiny\textsc{FSW, FPv}})}_{c,FRI}/2]<B<B_{c,FRI}$, and 
the magnetization is higher than $S-s$. The magnetization increases through the filling of this band, in accord 
to the Fermi distribution, up to the saturation value $m=s+S$, which is exponentially reached.

\subsection{Spin-wave theory - fully polarized vacuum}
\label{Section_FPv}
\begin{figure}[!htb]
\includegraphics[width=0.48\textwidth]{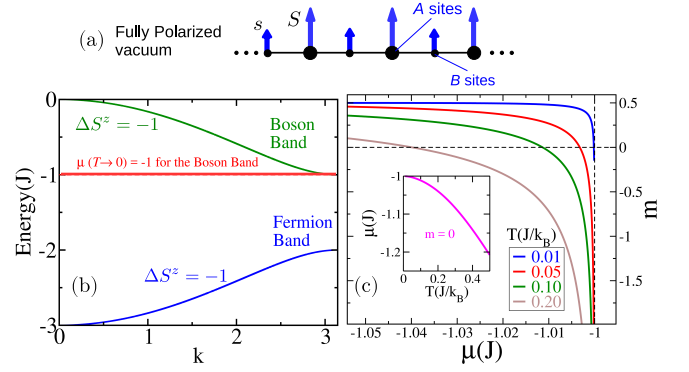}
\caption{(color online). Free spin-wave magnon branches from the classical ferromagnetic vacuum - calculating the thermodynamic properties.
(a) The classical fully polarized vacuum of the ($s$,$S$) chain.
(b) Free spin-wave (FSW) results for the magnon energies relative to the
fully polarized vacuum (FPv) for $T\neq0$ and $B=0$ for the ($s=1/2$, $S=1$) chain. 
In this case, both branches are ferromagnetic with magnons carrying a spin $\Delta S^z=-1$. 
To calculate the thermodynamic functions, the lower (higher) magnon band is filled following the Fermi (Bose) distribution function.
An effective chemical potential $\mu$ is introduced in the Bose distribution to prevent particle 
condensation at the $k=\pi$ mode for $B=0$ and $T\rightarrow0$.
The critical fields are $B^{(\textit{\tiny\textsc{FSW, FPv}})}_{c,FRI}=2.00J$ and $B^{(\textit{\tiny\textsc{FSW, FPv}})}_{c,FP}=3.00J$.
(c) The chemical potential $\mu$ is chosen such that $m=0$ for $B=0$.  
The inset shows that $\mu(T\rightarrow0)\rightarrow-1$ as $T\rightarrow0$. In this limit
only the lower energy band is occupied, implying that 
$m\rightarrow (S-s) =1/2$, the ferrimagnetic magnetization, as $T\rightarrow0$ and $B\rightarrow0$.}
\label{fig:swfpt}
\end{figure}

In this section, we study the free spin wave theory from a fully polarized vacuum, 
illustrated in Fig. \ref{fig:swfpt}(a). 
We show that this theory provides a good description of the low-$T$ physics, 
and is quantitatively much better than the free spin wave description from the ferrimagnetic 
vacuum. The critical saturation field has an exact value, while the critical 
field at the end of the ferrimagnetic plateau is $B^{(\textit{\tiny\textsc{FSW, FPv}})}_{c,FRI}=2J$.

The Holstein-Primakoff transformation in this case is
\begin{eqnarray}
S^+_j &=&\sqrt{2S}\Big(1-\frac{a^\dagger_j a_j}{2S}\Big)^{1/2}a_j\text{, and }S^z_j=S-a^\dagger_j a_j;\\ 
s^+_j &=&\sqrt{2s}\Big(1-\frac{b^\dagger_j b_j}{2s}\Big)^{1/2}b_j\text{, and }s^z_j=s-b^\dagger_j b_j,
\label{eq:hp-relations-fp}
\end{eqnarray}
with the two bosons lowering the site magnetization by one unit. To quadratic order in these bosonic operators, the Hamiltonian of the system, Eq. (\ref{HeisFerri}), 
is
\begin{eqnarray}
 \mathcal{H}^{\text{\tiny\textsc{(FSW-FPv)}}} & = & E^{\text{\tiny\textsc{(FPv)}}}_{class}+J\sum_{j}\Bigg\{-s\Big(a_{j}^\dag a_{j} + a_{j+1}^\dag a_{j+1}\Big) \nonumber \\ 
               & - & 2Sb_{j}^\dag b_{j}+\sqrt{sS}\Bigg[\Big(a_{j} + a_{j+1}\Big)b_{j}^\dag + \Big(a_{j}^\dag + a_{j+1}^\dag \Big)b_{j}\Bigg] \nonumber \\  
               & + & B\sum_j\Big(a_{j}^\dag a_{j} + b_{j}^\dag b_{j}\Big) \Bigg\},
\label{eq:h-fsw-fp}
\end{eqnarray}
with $E^{\text{\tiny\textsc{(FPv)}}}_{class} = 2JNsS -B\big(S+s\big)N$. Fourier transforming the bosonic operators and using the Bogoliubov transformation
\begin{eqnarray}
 a_{k}^\dag &=& \alpha_{k}^\dag\cos\theta_{k} - \beta_{k}^\dag\sin\theta_{k};\\
 b_{k}^\dag &=& \beta_{k}^\dag\cos\theta_{k} + \alpha_{k}^\dag\sin\theta_{k},
\end{eqnarray}
with
\begin{equation}
 \tan2\theta_k = 2\frac{\sqrt{sS}}{S-s}\cos \Big(\frac{k}{2}\Big),
\end{equation}
the Hamiltonian in Eq. \ref{eq:h-fsw-fp} is written as
\begin{equation}
 \mathcal{H}^{\text{\tiny\textsc{(FSW-FPv)}}}= E^{\text{\tiny\textsc{(FPv)}}}_{class} + \sum_{k}\Big[\omega^{\text{\tiny\textsc{(FPv)}}}_{k,1}\alpha_{k}^\dag \alpha_{k} + \omega^{\text{\tiny\textsc{(FPv)}}}_{k,0}\beta_{k}^\dag \beta_{k}\Big],
 \label{HamiltonianoGeralLivre}
\end{equation}
where the dispersion relations \cite{Maisinger1998} $\omega^{\text{\tiny\textsc{(FPv)}}}_{k,\eta}$ are
\begin{eqnarray}
 \omega^{\text{\tiny\textsc{(FPv)}}}_{k,\eta} &=& (-1)^{\eta+1}\sqrt{\big(S-s\big)^2 + 4sS{\cos}^2 \Big(\frac{k}{2}\Big)}\nonumber\\
       & &-\big(S + s\big)+B,
\label{eta0}
\end{eqnarray}
with $\eta=0\text{ or }1$.

To discuss the $T=0$ magnetization curve implied by these spin-wave modes, we present in Fig. \ref{fig:swfpt}(b) the dispersion relations 
$\omega^{\text{\tiny\textsc{(FPv)}}}_{k,\eta}$ for the ($s=1/2$, $S=1$) chain 
and $B=B^{(\textit{\tiny\textsc{FSW, FPv}})}_{c,FP}=2J(s+S)=3J$. 
At $B=B^{(\textit{\tiny\textsc{FSW, FPv}})}_{c,FP}=B_{c,FP}$, 
both bands are empty, and the magnetization is the fully polarized one.
Decreasing $B$, the $\eta=0$ band is filled in accord to Fermi-Dirac statistics, and 
the magnetization decreases. The critical field at the end point of the ferrimagnetic 
plateau is obtained making $\omega^{\text{\tiny\textsc{(FPv)}}}_{\pi,0}=0$, which implies 
$B^{(\textit{\tiny\textsc{FSW, FPv}})}_{c,FRI}=2SJ$, equal to $2J$ for the ($s=1/2$, $S=1$) chain. 
At this value of $B$, the $\eta=0$ 
band is totally filled and $m=(s+S)-1$, giving $1/2$ for the ($s=1/2$, $S=1$) chain. 
There is a gap of $2(S-s)J$ between the $\eta=0$ and $\eta=1$ bands, at $k=\pi$;
hence, the bosonic $\eta=1$ band should start to be filled at $B=B^{(\textit{\tiny\textsc{FSW, FPv}})}_{c,FRI}-2(S-s)J$, 
and the theory does not qualitatively reproduce the $T\rightarrow0$ magnetization curve.
This problem is overcome by considering the finite temperature theory, 
with Takahashi's constraint and effective chemical potential.  For finite $T$, the 
magnetization is given by
\begin{equation}
 m(T,B)= (S + s)-\frac{1}{N}\sum_{k}[n^{\text{\tiny\textsc{(FPv)}}}_{k,0}+n^{\text{\tiny\textsc{(FPv)}}}_{k,1}],
 \label{contraint-fp}
\end{equation}
where
\begin{eqnarray}
  n^{\text{\tiny\textsc{(FPv)}}}_{k,0} &=& \frac{1}{e^{\beta \omega^{\text{\tiny\textsc{(FPv)}}}_{k,0}} +1},\\
  n^{\text{\tiny\textsc{(FPv)}}}_{k,1} &=& \frac{1}{e^{\beta[\omega^{\text{\tiny\textsc{(FPv)}}}_{k,1}-\mu]} -1}.
\label{eq:n-fpv}
\end{eqnarray}
The constraint, which is applied at $B=0$, is
\begin{equation}
m(T,B=0)=0. 
\end{equation}
In Fig. \ref{fig:swfpt}(c) we present the magnetization as a function of the effective chemical $\mu$ for 
the indicated values of temperature. We note that $m\rightarrow-\infty$ as the temperature increases, 
similarly to the spin-wave theory with the ferrimagnetic vacuum. However, in this case $\mu\rightarrow-1$ 
as $T\rightarrow0$, as shown in Fig. \ref{fig:swfpt}(b). Hence, a finite chemical potential $\mu=-1$ 
associated to the bosonic $\eta=1$ band must be considered in the $T=0$ theory.
With this chemical potential, the $\eta=1$ band stays empty at $T=0$ for any value of $B$.   

The thermodynamic functions are calculated using Eq. \ref{eq:n-fpv}, with $\mu(T,B)=\mu(T,B=0)$. 
For finite $T$, the fermionic $\eta=0$ band is completely filled and the occupation of the $\eta=1$ band 
is such that $m=0$. Considering the low-$T$ regime, as $B$ increases, the energy of the two bands raises, lowering the total occupation of
the $\eta=1$ band, since $\omega^{\text{\tiny\textsc{(FPv)}}}_{k,1}-\mu$ linearly increases with $B$ for any $k$, 
and $m$ increases. The magnetization exponentially reaches its value at the ferrimagnetic plateau, $m=S-s$, as 
$B$ increases, since $n^{\text{\tiny\textsc{(FPv)}}}_{k,1}\rightarrow0$ for any $k$ and the $\eta=0$ band is completely 
filled. For $[B^{(\textit{\tiny\textsc{FSW, FPv}})}_{c,FRI}/2]<B<B^{(\textit{\tiny\textsc{FSW, FPv}})}_{c,FRI}$, with 
$[B^{(\textit{\tiny\textsc{FSW, FPv}})}_{c,FRI}/2]$ related to the point $B=B_{c,FRI}/2$ in Fig. \ref{fig:mb0-01}, the occupation 
of the $\eta=0$ band decreases from the $T=0$ case: $n^{\text{\tiny\textsc{(FPv)}}}_{k,0}=1$ for any $k$, and 
the magnetization is higher than $S-s$. 
The magnetization increases with $B$, and exponentially reaches the fully polarized value at 
$B>B^{(\textit{\tiny\textsc{FSW, FPv}})}_{c,FP}$, since magnons at the $\eta=0$ band are thermally excited.
\begin{figure}[!htb]
\includegraphics[width=0.36\textwidth]{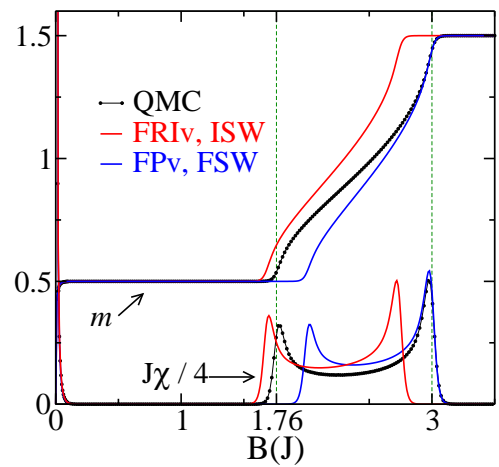}
\caption{(color online). Comparison between results from quantum Monte Carlo (QMC) method, $N=256$ unit cells, and the two spin-wave approaches for 
the magnetization per cell $m$ and the susceptibility $\chi$:
($s=1/2$, $S=1$) chain at temperature $T=0.02(J/k_B)$. Results from the interacting spin-wave theory from a ferrimagnetic vacuum (ISW-FRIv) 
and free spin-wave theory from a ferromagnetic vacuum (FSW-FPv) compare well with
QMC for $B\lesssim B_{c,FRI}$ and $B\gtrsim B_{c,FP}$. The maximum in $\chi$ related to $B_{c,FRI}$ ($B_{c,FP}$) 
is better localized, compared to QMC, through the ISW-FRIv (FSW-FPv) approach.}
\label{fig:comp}
\end{figure}

\subsection{Comparison between QMC data and the two spin-wave approaches}

In Fig. \ref{fig:comp} we present magnetization and susceptibility $\chi=\partial m/\partial B$ as a function 
of $B$ from ISW-FRIv and FSW-FPv theories along with QMC data, at $T=0.1J$. Since the ISW-FRIv gives a better 
result for $B_{c,FRI}$, this theory is better in the vicinity of this critical field. Otherwise, the 
FSW-FPv approach is better in the vicinity of $B_{c,FP}$. Further, the amplitudes of the two peaks
in $\chi(B)$, which marks the crossover to the LL regime, have values lower than the ones 
given by QMC. The difference 
between the amplitudes of the spin-wave approaches and QMC data is related
to limitations in the spin-wave theories.
Despite it, the description from both spin-wave theories are qualitatively excellent, and 
quantitatively very acceptable in the low-$T$ regime.

Below we calculate the $T$ vs $B$ phase diagram in the low-$T$ regime from the FSW-FPv theory. We  
study the crossover lines between the LL regimes and the quantum critical regimes; as well as the crossovers 
lines between the plateau regimes and the quantum critical regimes. 
We use the FSW-FPv approach since it has essentially the same precision of 
the ISW-FRIv theory, if we consider a range of $B$ from 0 to the saturation field; also, the critical point 
$B_{c,FP}$ is exact in the FSW-FPv theory. 

\section{Luttinger liquid regime}
\label{sec:ll-regime}

In the LL phase, the dispersion relation can be approximated by $\pm v_F|k-k_F|$, where $v_F$ is the Fermi velocity.
Further, in this regime the magnetization has the 
form \cite{PhysRevLett.99.057205}:
\begin{equation}
 m=m(T=0)-\frac{\pi}{6v_F^2}\frac{\partial v_F}{\partial B}(k_BT)^2+O(T^3).
 \label{eq:m-cft}
\end{equation}
In our case, the Fermi velocity along the $\eta=0$ band is $v_F=[\partial\omega^{\text{\tiny\textsc{(FPv)}}}_{k,0}/\partial k]_{k=k_F}$, 
with $k_F$ calculated from $\omega^{\text{\tiny\textsc{(FPv)}}}_{k,0}\vert_{k=k_F}=0$.  

\begin{figure}[!htb]
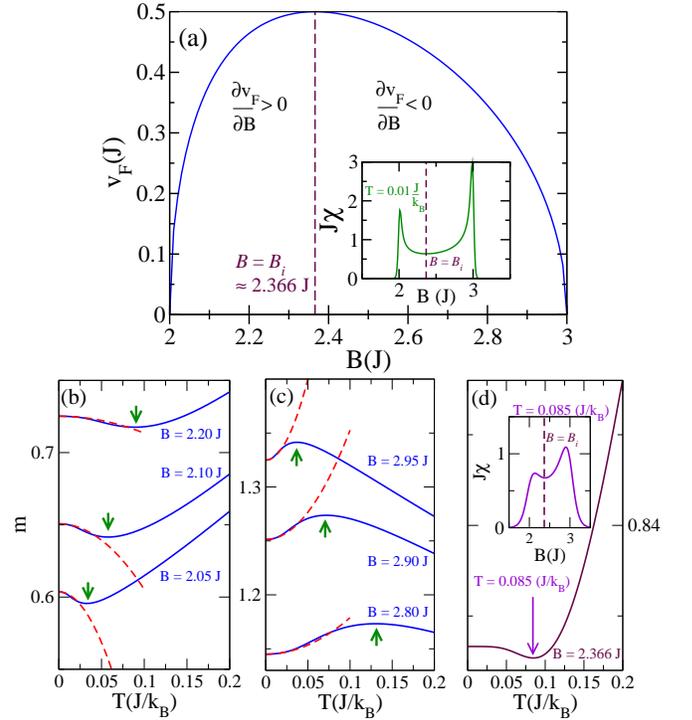

\includegraphics*[width=0.35\textwidth]{fig5a.eps}
\includegraphics*[width=0.48\textwidth]{fig5bcd.eps}
\caption{(color online). Results from the free spin-wave approach with the fully polarized vacuum (FSW-FPv).
(a) Fermi velocity $v_F$ as a function of the 
magnetic field $B$ and [(b), (c) and (d)] 
magnetization curves $m(T)$. (a) $\partial v_F/\partial B\rightarrow+\infty$ and $v_F\rightarrow0$ as $B\rightarrow\bcfsw{FRI}=2.00J$, 
while $\partial v_F/\partial B\rightarrow-\infty$ and $v_F\rightarrow0$ as $B\rightarrow\bcfsw{FP}=3.00J$.
As shown in the inset, for $B=B_i\approx2.366J$, 
$\partial v_F/\partial B=0$ and the susceptibility $\chi(B)$ has a minimum at this value of $B$. 
(b) $m(T)$ for the indicated values of $B$ in the vicinity of the critical field $\bcfsw{FRI}$.
(c) $m(T)$ for values of $B$ in the vicinity of the critical field $\bcfsw{FP}$.
(d) $m(T)$ for $B=B_i$. The $m(T)$ curves to order $O(T^2)$, Eq. (\ref{eq:m-cft}),  
are shown as dashed lines in (b) and (c) for the corresponding values of $B$, arrows indicate 
local extreme points in $m(T)$, which are used as a criterium to identify the LL regime. The inset in (d) shows that 
the minimum in $m(T)$ is associated to the local minimum in $\chi(B)$, which is found between the two critical fields.
}
\label{fig:mtfp}
\end{figure}

In Fig. \ref{fig:mtfp}(a) we present $v_F$ as a function of $B$ for the (1/2,1) chain.
Near the critical fields, $|\partial v_F/\partial B|$ is large and $v_F$ little. 
For a fixed $B\gtrsim B^{(\textit{\tiny\textsc{FSW, FPv}})}_{c,FRI}$, as shown in Fig. \ref{fig:mtfp}(b), 
the magnetization presents a fast decay from the $T=0$ value as $T$ increases.
Also, for $B\lesssim B^{(\textit{\tiny\textsc{FSW, FPv}})}_{c,FP}$, as shown in Figs. \ref{fig:mtfp}(c),
$m$ increases from $m(0)$. In both cases, the curvature of the $m(T\rightarrow0)$-curve increases as 
$B$ get closer to the critical fields. 
The crossover temperature $T(B)$ of the LL regime at a fixed $B$ is 
defined as the point at which $m(T)$ departs from the quadratic behavior
in Eq. (\ref{eq:m-cft}).
So, $T(B)$ is taken to be at the minima ($B\gtrsim B^{(\textit{\tiny\textsc{FSW, FPv}})}_{c,FRI}$) and 
maxima ($B\lesssim B^{(\textit{\tiny\textsc{FSW, FPv}})}_{c,FP}$) of the $m(T)$ curve \cite{PhysRevLett.99.057205}.
In particular, as $B\rightarrow B_c$ the crossover line separates the LL regime and the 
quantum critical regime, for which the excitations have a quadratic dispersion relation. 
In this case, a universal, model independent, 
straight line $k_BT(B)=a|B-B_c|$, with $a=0.76238$, 
can be derived \cite{PhysRevLett.99.057205}.
\begin{figure}[!htb]
\includegraphics*[width=0.44\textwidth]{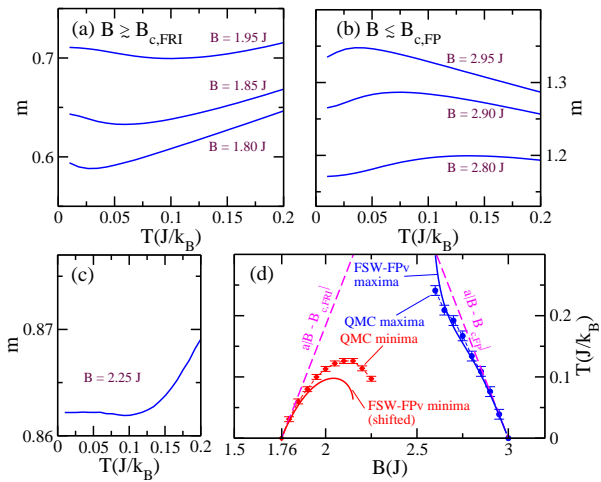}
\caption{(color online). Magnetization per cell $m(T)$ with fixed $B$: calculating the crossover lines bounding the Luttinger liquid regime.
Quantum Monte Carlo (QMC) results for the magnetization curves $m(T)$ and 
the crossover lines for a system with $N=128$. (a) $m(T)$ for  
values of $B$ in the vicinity of the critical field $B_{c,FRI}=1.76J$. 
(b) $m(T)$ for values of $B$ in the vicinity of the critical field $B_{c,FP}=3.00J$.
(c) $m(T)$ for a value of $B$ such that $\partial \chi/\partial B\approx0$ at $T=0$ and inside the Luttinger liquid phase, 
dashed line in Fig. \ref{fig:mb0-01}. (d) Local extreme points of $m(T)$ curves from QMC and free spin-wave from the 
fully polarized vacuum (FSW-FPv). In the case of the FSW-FPv local minima, we shift $B$ by $B_{c,FRI}-\bcfsw{FRI}\approx0.24J$. 
The exact crossover straight lines as $T\rightarrow 0$, extended in the figure for better visualization: $a|B-B_{c,FRI}|$ and $a|B-B_{c,FP}|$, 
  with $a=0.76238$, are also shown. The error bars are defined as half the temperature step ($\Delta T=0.008$) used 
  to calculate $m(T)$.}
\label{fig:mtQMC}
\end{figure}

In the inset of Fig. \ref{fig:mtfp}(a), we show that the minimum in the $\chi(B)=\partial m/\partial B$ curve 
is found at $B=B_i$, a value of $B$ at which $|\partial v_F/\partial B|=0$.
This value of $B$ marks a crossover from the regime 
where excitations are predominantly from the FRI critical state to the
regime where they come from the FP critical state.
At $B=B_i$, the Fermi wave-vector is at the inflection point of the dispersion curve 
($d^2 \omega^{\text{\tiny\textsc{(FPv)}}}_{k,0}/dk^2=0$), since
\begin{equation}
 \frac{\partial v_F}{\partial B}=\left[ \frac{d^2\omega^{\text{\tiny\textsc{(FPv)}}}_{k,0}}{dk^2}\right ]_{k=k_F}\left (\frac{\partial k_F}{\partial B} \right), 
\end{equation}
and $k_F$ increases monotonically with $B$ between the critical fields.
If the value of $k$ at the inflection point is $k_i$, we can calculate 
$B_i$ from the equation $\omega^{\text{\tiny\textsc{(FPv)}}}_{k_i,0}=0$.
For the (1/2,1) chain, for example, $B_i=2.366J$ and is indicated in Fig. \ref{fig:mtfp}(a). 

At $B=B_i$, $\partial v_F/\partial B=0$ and the quadratic term 
in Eq. (\ref{eq:m-cft}) is absent. So, the more stable, against $T$, LL 
region is found for $B\approx B_i$. Since the crossover temperatures $T(B)\rightarrow 0$ 
near the critical fields, the $T(B)$ line has an \textit{asymmetric dome-like} profile, which 
is a consequence of the $v_F$ curve, shown in Fig. \ref{fig:mtfp}(a) for the case of the (1/2,1) chain,
and is also observed in other quantum magnets \cite{Zapf2014}.

A minimum in the $m(T)$ curve is 
also observed for $B=B_i$, due to the $O(T^3)$ in Eq. (\ref{eq:m-cft}), as shown in Fig. \ref{fig:mtfp}(d).
In this case, however, this extreme point is associated with the 
minimum in the $\chi(B)$ curve, at $B=B_i$, as shown in the inset of 
Fig. \ref{fig:mtfp}(d).

In Fig. \ref{fig:mtQMC} we show $m(T)$ curves for the (1/2,1) chain calculated with QMC method 
to discuss the qualitatively
agreement between these almost exact results and the conclusions from the spin-wave theory.
In Figs. \ref{fig:mtQMC}(a) and (b), we show the minimum (maximum) in the $m(T)$ curve for 
$B\gtrsim B_{c,FRI}=1.76J$ ($B\lesssim B_{c,FP}=3J$). In Fig. \ref{fig:mtQMC}(c), we calculate 
$m(T)$ for a value of $B$ in the vicinity of the minimum in the $\chi(B)$ curve, $B=B_i$.
Using the data in Fig. \ref{fig:mb0-01}, it is located at $B_i=(2.27\pm0.07)J$, and is indicated as a 
dashed line in that figure. As shown in Fig. \ref{fig:mtQMC}(c), the $m(T\rightarrow0)$ curve is also 
flat, as in Fig. \ref{fig:mtfp}(d), for $B=2.25J$. The minimum in the $m(T)$ curve appears at 
$T\approx0.1J$. As can be observed in the $T=0.1J$ susceptibility curve in Fig. \ref{fig:mb0-01}, it 
is also associated with the minimum in the $\chi(B)$ curve, at $B\approx B_i$. 

In Fig. \ref{fig:mtQMC}(d), we compare the position of the local extreme points in the $m(T)$ curves
from QMC and FSW-FPv methods. The values of $B$ at the minima of $m(T)$ were translated by  
$B_{c,FRI}-\bcfsw{FRI}\approx0.24J$. The lines for the maxima in $m(T)$ from both 
methods are in very good agreement since the FSW-FPv is almost exact for $T\rightarrow0$, due 
to the low density of excited magnons in this temperature regime. Otherwise, the minima 
from both methods do not compare well, except for $T\rightarrow0$, which is dominated by the critical point.
\begin{figure}[!htb]
\begin{center}
\includegraphics[width=0.4\textwidth]{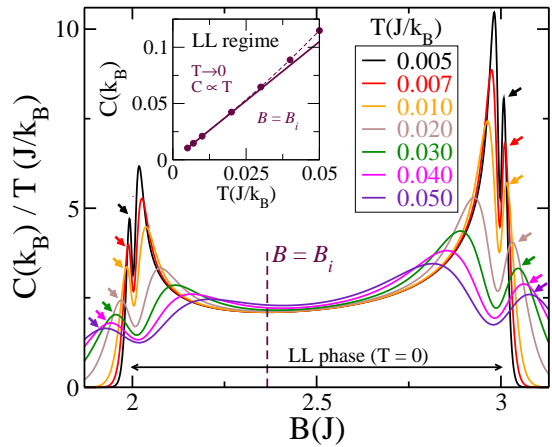}
\caption{(color online). Specific heat from the free spin-wave theory from a fully polarized vacuum (FSW-FPv)  
for $T\rightarrow0$. In the Luttinger liquid (LL) regime, $C\sim T$ as $T\rightarrow0$, and $C/T$ is approximately constant
for $B\approx B_i=2.366J$. The inset shows this linear behavior of $C$ at $B=B_i$.
The crossover from the $T=0$ insulating
plateau regime to the gapless quantum critical regime, at local maxima, are indicated by arrows.}
\label{fig:ct}
\end{center}
\end{figure}

We determine the crossover lines between the LL and plateau regimes through specific heat data, $C(B)$.
In Fig. \ref{fig:ct} we present FSW-FPv results for $C(B)$ 
in the low-$T$ regime. In the LL phase, at $T=0$, the specific heat $C\sim T$ as $T\rightarrow0$, 
and $C/T$ is approximately constant in the LL regime, as shown in Fig. \ref{fig:ct}. 
The range of $B$ near $B= B_i$ is the more robust for this regime, and we present in the inset of Fig. \ref{fig:ct} 
the linear behavior of $C$ as a function of $T$.
For $B\lesssim\bcfsw{FRI}$ or $B\gtrsim\bcfsw{FP}$, the excitations are exponentially activated
and the crossover to the quantum critical regime is marked by a local maximum in $C(B)$. 
The points of these crossover lines, $T_{\text{plateau}}(B)\sim|B-B_c|$, are indicated by arrows in
Fig. \ref{fig:ct}. The quantum critical regime is bounded by this crossover line and that of the 
LL regime, which points appears as a second local maximum near $\bcfsw{FRI}$ and $\bcfsw{FP}$
in Fig. \ref{fig:ct}. 
\section{Summary and discussions}
\label{sec:summary-pd}
\begin{figure}[!htb]
\begin{center}
  \includegraphics[width=0.47\textwidth]{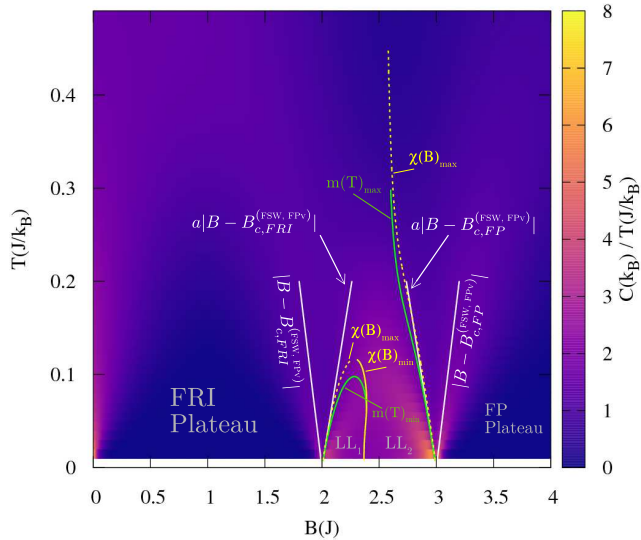}
  \caption{(color online). Spin-wave $T - B$ phase diagram of the ($s=1/2$, $S=1$) chain from the FPv. 
  The quantum critical points $B^{(\textit{\tiny\textsc{FSW, FPv}})}_{c,FRI}=2.00J$ and 
  $\bcfsw{FP}=3.00J$ bound the FRI and FP plateau regions, respectively. 
  Increasing temperature, the plateau width decreases and the lines $k_BT=|B-\bcfsw{FRI}|$ and $k_BT=|B-\bcfsw{FP}|$ limit the 
  plateau regions for $B\lesssim B_c$ [ferrimagnetic (FRI) plateau] and $B\gtrsim B_c$ [fully polarized (FP) plateau]. The LL regime has crossover lines given by $a|B-\bcfsw{FRI}|$ and $a|B-\bcfsw{FP}|$, 
  with $a=0.76238$, for $B\rightarrow B_c$, as indicated by local maxima of the susceptibility 
  $\chi(B)=\frac{\partial m}{\partial B}$, $\chi(B)_{max}$. Between these local maxima, there is a local minimum [$\chi(B)_{min}$] 
  separating the regions under the influence of the $\bcfsw{FRI}$ critical point and that of the $\bcfsw{FP}$ one.}
  \label{fig:pd}
\end{center}
\end{figure}

We have calculated the critical properties of alternating ferrimagnetic chains in the presence of a magnetic field from two
spin-wave theories. We determine the better low-energy description of the excitations, considering the level of approximation,
comparing the results with quantum Monte Carlo data. These ferrimagnetic chains present 
two magnetization ($m$) plateaus, the ferrimagnetic (FRI) plateau, for which $m=S-s$ and the fully polarized (FP) one, at $m=s+S$.
The first spin-wave theory, is an interacting spin-wave (ISW) approach with the FRI classical vacuum, ISW-FRIv. The second methodology, 
is a free spin-wave (FSW) calculation from the FP state, FSW-FPv. In both cases, two bands are obtained. To calculate the finite temperature ($T$) 
properties of the system, one of the bands is considered as a bosonic band, with an effective chemical potential to 
prevent boson condensation at $B=0$; while the other is considered as a hard-core boson band, with a fermionic one-particle 
thermal distribution. Near the endpoint of the FRI plateau, the ISW-FRIv theory is a better option; while 
the FSW-FPv is exact for $T\rightarrow0$ 
near the endpoint of the FP plateau. Since we are interested in describing the whole $T$ vs. $B$ phase diagram of 
the system, we deepen the study on the FSW-FPv, calculating the finite $T$ crossover lines bounding the plateau and the Luttinger liquid (LL) regimes.

In Fig. \ref{fig:pd} we summarize our results in a $T$ vs. $B$ phase diagram, and show specific heat data $C/T$ as a function of
$B$ and $T$. In the FRI and FP plateau regions the excitations are gapped, and $(C/T)\rightarrow0$ as $T\rightarrow0$. The 
gaps close at the quantum critical (QC) fields $\bcfsw{FRI}=2J$ and $\bcfsw{FP}=3J$, and local maxima appears in the values of 
$C/T$ for a fixed $T$. These local maxima indicate the crossover between the plateau and the QC regimes, and 
between the QC and LL regimes. As $T\rightarrow0$, the crossover line between the plateau and the QC regimes (P-QC line) is
a straight line $k_BT(B)=|B-B_c|$, 
for $B_c=\bcfsw{FRI}$ and $B_c=\bcfsw{FP}$; while a straight line $a|B-B_c|$, with a model-independent constant $a=0.76238$, 
marks the crossover between LL and QC regimes (LL-QC lines). The LL-QC line which contains the critical point $B=\bcfsw{FRI}$ $[B=\bcfsw{FP}]$ 
was also calculated from local minima (local maxima) in
the $m(T)$ curves: $m(T)_{min}$ $[m(T)_{max}]$. The LL-QC lines were also calculated from local maxima in the susceptibility curve 
$\chi(B)$ at fixed $T$: $\chi_{max}(B)$. 

The Luttinger liquid regime can 
be divided into two regions, separated by the minimum in the $\chi(B)$ curve with a fixed temperature, $\chi_{min}(B)$. 
The value of the magnetic field at which this minimum occurs at $T=0$, $B_i$, is at 
the inflection point of the magnon band and changes little with $T$. The line $m(T)_{min}$ as a function of $B$ 
meets the line $\chi_{min}(B)$ for $B\approx B_i$. Finally, the LL regime has an asymmetric dome-like profile which is associated with 
the Fermi velocity profile as a function of $B$ at the relevant magnon band, as observed in other quantum magnets \cite{Zapf2014}.

We acknowledge financial support from Coordenação de Aperfeiçoamento de Pessoal de Nível Superior (CAPES), 
Conselho Nacional de Desenvolvimento Cientifico e Tecnológico (CNPq), and Fundação de Amparo à Ciência e Tecnologia de Pernambuco (FACEPE), 
Brazilian agencies, including the PRONEX Program of FACEPE/CNPq.

\end{document}